\documentstyle[11pt,epsf]{article}
\newcommand{\beq}{\begin{equation}}
\newcommand{\eeq}{\end{equation}}
\newcommand{\beqa}{\begin{eqnarray}}
\newcommand{\eeqa}{\end{eqnarray}}
\newcommand{\beqan}{\begin{eqnarray*}}
\newcommand{\eeqan}{\end{eqnarray*}}

\newcommand{\ben}{\begin{enumerate}}
\newcommand{\een}{\end{enumerate}}
\newcommand{\bfl}{\begin{flushleft}}
\newcommand{\efl}{\end{flushleft}}
\newcommand{\ba}{\begin{array}}
\newcommand{\ea}{\end{array}}
\newcommand{\btab}{\begin{tabular}}
\newcommand{\etab}{\end{tabular}}
\newcommand{\bit}{\begin{itemize}}
\newcommand{\eit}{\end{itemize}}

\newcommand{\hs}{\hspace}

\def \g {\gamma}
\def \g5 {\gamma_{5}}
\newcommand{\prepr}[1] {\begin{flushright} {\bf #1} \end{flushright} \vskip
1.5cm}
\newcommand{\titul}[1] {\begin{center}{\Large {\bf #1 } } \end{center} \vskip 1.cm}

\newcounter{muni}

\begin{document}
\vspace{.1cm}
\hbadness=10000
\pagenumbering{arabic}
\begin{titlepage}
\prepr{Preprint hep-ph/9601226\\PAR/LPTHE/96-01 }
\titul{A new CP violation mechanism generated by the standard
neutral Higgs boson :
the $\eta  \rightarrow \pi + \pi $ case.}
\vspace{5mm}
\begin{center}

{\bf X.Y.Pham \footnote{\rm Postal address: LPTHE, 
Tour 16, $1^{er}$ Etage, 
4 Place Jussieu, F-75252 Paris CEDEX 05, France. \\
. \hspace{5mm} Electronic address : pham@lpthe.jussieu.fr }
and M. Gourdin \footnote{\rm Postal address: LPTHE, 
Tour 16, $1^{er}$ Etage, 
4 Place Jussieu, F-75252 Paris CEDEX 05, France. \\ 
. \hspace{5mm} Electronic address : gourdin
@lpthe.jussieu.fr } }
\end{center} 

\vspace{15mm}
\begin{center}
{\large \bf \it
Universit\'e Pierre {\it \&} Marie Curie, Paris VI \\
Universit\'e Denis Diderot, Paris VII \\
Physique Th\'eorique et Hautes Energies
} 
\end{center}

\thispagestyle{empty}
\newpage
\hspace{1cm} \large{} {\bf  \hspace{3mm} Abstract}    
\vspace{0.5cm}
\normalsize

Strictly within the standard electro-weak interaction, CP violation 
in the flavour conserving process $\eta \rightarrow \pi + \pi $ could
originate from the mixing of the $\eta$ meson with the virtual scalar 
Higgs boson $ H^{0}$ via $W$ and top quark exchange.

The parity-violation carried by  weak gauge bosons makes the mixing
possible by quantum effect at two-loop level. 
Nowhere the Kobayashi-Maskawa (KM) phase mechanism is needed.
The phenomenon reveals an unexpected new role of the Higgs boson in the
CP symmetry breaking.

For the Higgs mass between 100-600 GeV,
the  $\eta \rightarrow \pi +\pi $
branching ratio is found to be $3.6 \cdot 10^{-26} - 2.4 \cdot 10^{-29}$ ,
hence CP violation mechanisms beyond the Standard  Model are the only
ones that could give rise to its observation at existing or near future
$\eta$ factories, unless the Higgs mass is improbably as light as 550 $MeV$.  

\vspace{10mm}
{\bf PACS numbers : 11.30.Er , 12.15.Ji , 13.25.Jx ,14.8o Bn ,11.40.Ha }
\end{titlepage}

\newpage

To understand the origin and the nature of CP violation, in addition to
the studies of flavour changing K and B mesons processes, investigations
are also needed in flavour conserving ones \cite{R1} for which the
$\eta \rightarrow \pi + \pi $ 
decay and the electric dipole moment of baryons are some typical examples. Like
the $K_{L}^{0}$, an eventual coexistence of both three and two-pion decay
modes of the $\eta$ would imply that CP is violated in the flavour
conserving sector. Therefore experimental searches for the two-pion decay
mode of the $\eta$ is of great interest \cite{R1}, and the purpose of this
letter is two fold : 
\begin{enumerate}
\item point out an unexpected new role of the Higgs boson in the CP breaking,
generated by quantum effect.
\item  give a reliable estimate of its branching ratio.
\end{enumerate}

We do find indeed that CP violation in  $\eta \rightarrow \pi +  \pi $ can be
triggered by the neutral Higgs boson, {\it independently of all
other mechanisms} \cite{R2, R3, R4, R5} :
the decay can occur {\it not by explicit
CP non conservation} term put by hand in the lagrangian
(like the standard KM non-zero phase assumed from the start),
but by a completely different mechanism through the mixing between the
$\eta$ and the neutral Higgs boson.

This point is basically new, it can only occur at two loop level as
shown in Figs.1,2,3 using the renormalizable gauge $R_{\xi}$ :
besides the $W^{\pm}, Z ^0$ gauge bosons, the would-be $\chi^{\pm}, \chi^3$
Goldstone-Higgs fields (those absorbed by $W^{\pm}, Z^0$ to get masses)
also contribute. They represent the complete set of two loop 
diagrams that participate to the $\eta$-Higgs mixing. At one loop level, 
the mixing cannot take place, there is no way 
to get rid of the $\gamma_5$ coupling of the $\eta$ with its
antisymmetric tensor $ i\varepsilon_{\mu\nu\rho\sigma}$, while the Higgs
coupling is symmetric.

The main reason for the mixing to occur is the following: the $\gamma_5$ 
coupling of the $\eta$ with quarks when combined with the product 
$V \times A$ of the gauge boson couplings will be absorbed and give rise to 
a symmetric tensor $g_{\mu\nu}$ and makes the mixing with the Higgs boson 
possible.

Since both the $\eta$ meson and the Higgs boson have $ C=+1$, physically it 
means that through two loop integrals, the parity-violation  
$V \times A$  property of the weak bosons can shift the intrinsic  P = $- 1$ 
of the  $\eta$ into the  P = $+ 1$ of the $H^{0}$, hence CP 
$ -1 \leftrightarrow +1$ mixing : {\it parity-violation turns out to be the source 
of CP non-conservation, due to the Higgs and gauge bosons interplay}. This 
observation is first illustrated by explicit computation of the Figure 1 diagram, 
the relevant quantity to be considered is:
\begin{equation}
I(k^2) \equiv (-1) \int {d^4 q \over (2\pi)^4} {d^4 p \over (2\pi)^4}
{ Trace [ \gamma_5 (\not{p} + m) \gamma_{\alpha} (1 -  \gamma_5 )
(\not{q} + m^{'}) \gamma_{\beta} (1 - \gamma_5 )
(\not{p} + \not{k} + m)] g^{\alpha\beta} \over
(q^2 - {m^{'}}^2)(p^2 - m^2)((p+k)^2 - m^2)((p-q)^2 - M^2)((p+k-q)^2 - M^2)}
\label{eq:1}
\end{equation}
with $ m = m_{s}, m^{'} = m_{c}, M = M_{W} $.
The contributions of other graphs will be given later.
In Eq.(1) we have taken, as an illustrative example, the
$s\overline{s}$ component of the $\eta$ in the loop. 

An important remark is in order :
Unlike the $W^{+} + W^{-}$ exchange in Fig. 1, for the $Z^0 + Z^0$ one, 
there are in fact two identical diagrams with fermion circulates 
in opposite directions around the loop,
their contributions give rise to only antisymmetric tensor 
and make the mixing with the Higgs boson impossible.
The reason is that $Z Z$ are identical particles
(similar to $\eta \rightarrow \gamma \gamma$ case).
However for the $W^{+} W^{-}$ case, there is only one diagram. 
To check this point one can go back to the $T$ products
$T(H_{\eta}(x) H_{W}(y) H_{W}(z))$,
apply the Wick theorem and rediscover the Feynman rules and graphs.
Here   $H_{\eta} = \eta \overline{s} \gamma_5 s$,
$H_{W} =  \overline{c} \gamma_{\mu} (1 - \gamma_5) s \hspace{2mm}W^{\mu} + h.c.$.
We can easily understand this point even with tree diagrams,
let us compare $s + \overline{s} \rightarrow Z^0 + Z^0$
to $s + \overline{s} \rightarrow W^{+} + W^{-}$.
  
To compute Eq.(1) we first integrate over $ d^4 p $  using the  $ x , y , z $  
Feynman parametrization, and then over $d^4 q $. Let us sketch out
the sucessive steps. After calculating the trace, 
we obtain:
\begin{equation}
I(k^2)  =  16m \int {d^4 q \over (2\pi)^4} { kq 
\over (q^2-{m^{'}}^2)} \,J(k,q)  
\label{eq:2}
\end{equation}
where :
\begin{eqnarray}
J(k,q) &=& \int {d^4 p \over (2\pi)^4} {1\over(p^2 - m^2)((p+k)^2 - m^2)
((p-q)^2 - M^2)((p+k-q)^2 - M^2)} \cr
&=& {i\pi^2\over(2\pi)^4} \int^1_0 dx \int ^x_0 dy \int ^y_0 {dz 
\over (x-z)^2(1-x+z)^2} {1\over (q^2+2qK -L^2)^2} 
\label{eq:3}
\end{eqnarray}
\begin{equation}
K= k {z(1-y)-y(1-x) \over(x-z)(1-x+z)} \hspace{5mm}, \hs{5mm} 
L^2 = {M^2(x-z) -m^2(1-x+z) + k^2y(1-y)\over (x-z)(1-x-z)},
\label{eq:4}
\end{equation}
When we neglect $m^2, {m^{'}}^2, k^2 $ compared to $M^2$, then we find:
\begin{equation}
\int {d^4 q \over (2\pi)^4} {kq \over (q^2-{m^{'}}^2)(q^2 + 2qK- L^2)^2 } =
i\pi^2 {k^2 \over 2M^2} {z(1-y)-y(1-x) \over x-z } + {\cal O} 
\left( {k^4 \over M^4}, 
{m^2 \over M^2}, {{m^{'}}^2 \over M^2} \right)
\label{eq:5}
\end{equation}
such that
\begin{equation}
I(k^2) =  C \hspace{2mm} {m  \over 32 \pi^{4} } \hspace{2mm} {k^{2} \over M^{2}} 
[ 1 + {\cal O} \left( {k^{2} \over M^{2}}, 
{m^{2} \over M^{2}}, {{m^{'}}^{2} \over M^{2} } \right) ] 
\label{eq:6}
\end{equation}
where 
\begin{equation}
C   \equiv \int ^1_0 dx \int ^x_0 dy \int ^y_0 dz  
{y(1-x)-z(1-y) \over (x-z)^{3} (1-x+z)^{2}}  =  {1 \over4}
\label{eq:7}
\end{equation}
The computation of C in Eq.(7) is tedious, numerical integration is helpless 
since there is a delicate cancellation of infinities. We have done it 
analytically by hand, step by step. The result is C=1/4.
The expression (6) we obtain for the two-loop integration 
is impressively simple, because higher orders in ${k^{2} \over M^{2}}$  , 
${m^{2}\over M^{2}}$ , ${{m^{'}}^{2} \over M^{2}}$ 
(beyound the linear term ${k^{2} \over M^{2}}$) are 
neglected in the course of our  $d^{4}q $  integration. 
In the  $d^{4}p$  one, everything is kept however. Without this legitimate 
approximation, we would obtain an avalanche of - unnecessary and numerically 
negligible- complicated expressions involving, among others, the dilogarithmic 
(or Spence) ${\cal L}_2$ function frequently met in such circumstance.

The diagram 1 looks like the familiar triangle in $\pi^0 \rightarrow 2\gamma $ 
with the external gauge bosons momenta integrated.
Immediately a question arises whether or not our { \it convergent and finite 
$I(k^2)$ term } has something
to do with the possible anomalous Ward identity \cite{R6} via 
$k^{\mu} P_{\mu}(k^2) \equiv Y(k^2)$ where $P_{\mu}(k^2)$ is defined similarly 
to $I(k^2)$ of Eq.(1) in which the first $ \gamma_5$ at the extreme left of the 
numerator of Eq.(1) is replaced by $\gamma_{\mu}\gamma_5$. 
The interrelation between $I(k^2)$ and $P_{\mu}(k^2)$ is respectively similar 
to that between the pseudotensors $R_{\alpha\beta}(k^2)$ and 
$T_{\mu\alpha\beta}(k^2)$ intervened in $\pi^0 \rightarrow  2\gamma $.
We find out, after the trace calculation, that $Y(k^2)$ could be written as
\begin{equation}
Y(k^2) = 16 \int {d^4q \over (2\pi)^4 }
{ F(q,k) \over q^2-m'^2}  
\label{eq:Y}
\end{equation}
where after the integration over $d^4p$, 
$F(q,k)$ is found to be 
\begin{equation}
F(q,k) = 
{i\pi^2\over(2\pi)^4} \int^1_0 dx \int ^x_0 dy \int ^y_0 {dz 
\over (x-z)^2(1-x+z)^2} {N(q,k) \over (q^2+2qK -L^2)^2}
\label{K}
\end{equation}
with
\begin{equation}
N(q,k) =  qk \{ -2q^2(x-z)^2 +2qk[z(1-y)-y(1-x)] - (M^2 -m^2)(x-z) + 2k^2y(1-y)\} 
-q^2k^2 (x-z)(1-2y)
\label{N}
\end{equation}
As expected, the integral over $d^4q$ in Eq. (8) is formally divergent, and 
in the regularization procedure one might think that if one tries to imitate 
the $\pi^0 \rightarrow  2\gamma $ case, then by a careless 
shift of the integration $p,q$ variables, one would rediscover an analogy with 
the false Sutherland-Veltman theorem \cite{R7} while a careful shift would 
lead to an analogy with the ABJ anomaly \cite{R6}. However the 
analogy stops here, since contrary to the $\pi^0 \rightarrow  2\gamma $ case 
for which there is always one $\gamma_5$ with the tensor 
$\varepsilon_{\mu\nu\alpha\beta}$ source of the ABJ anomaly, here in our case 
{\it the  $\gamma_5$ does not intervene}, the quantities $P_{\mu}(k^2)$ and 
$ I(k^2)$ are respectively vectors and scalar objects, contrary to the
corresponding pseudotensors $ T_{\mu\alpha\beta}(k^2)$ and 
$R_{\alpha\beta}(k^2)$ in the  $\pi^0 \rightarrow  2\gamma $ case, from that 
chiral anomaly was discovered. It is interesting to note that when we cut the
diagram 1 at the two $W^+, W^-$ lines, i.e. we compute the one loop triangle
$\eta \rightarrow W_{\alpha}^{+} +W_{\beta}^{-}$, we get the imaginary 
part Im $I(k^2)= 0$, thus checking by a dispersion relation that 
$I(k^2)$ is real, as we have already obtained before in Eq.(6).

We now go on  to the other diagrams.
For the diagram 2, the calculation is more complicated than the diagram 1,
and we get result similar to Eq.(6) with $C$ replaced by 
\begin{equation}
D_t \equiv \rho ( 1 + {\rho \over 2} ) \vert V_{ts} \vert^2 D(\rho)
\end{equation}
where
\begin{eqnarray}
D(\rho) &=& {1 \over (\rho - 1)^2} + {\rho \over 4 (\rho - 1)^3  }
+ {\rho (2 \rho - 1) \over 2 (\rho -1)^3} \log\rho \cr
&+&{\rho^2 \over (\rho - 1)^2} \left[ {\cal L}_2({\rho -1 \over \rho}) - 
{\pi^2 \over 6} \right]
+ {\rho \over (\rho -1)^3} \int^{1}_{0} dx (1 - 2x) {\cal L}_2 
( {x + \rho -1 \over \rho} ).
\end{eqnarray}
and $\rho = m_t^2/M_W^2$.

The latter integration can be also analytically computed in terms of products 
of the logarithmic function, its expression is rather combersome and not given
here. Also we have computed all other contributions denoted collectively by E.
Numerically, it turns out that the diagram 1 is dominant.\footnote
{The details of calculations will be given elsewhere, contributions of all 
other two loop diagrams are negligible: the up, top quark contributions in 
Fig. 1, the up, charm quark as well as the $Z^0, \chi^3$ in Fig. 2. As for 
the diagram 3, it is even smaller.}

It remains two questions to be settled: the first one concerns the effective
point-like coupling constant $ g_{\eta Q \overline{Q}} $ of the $\eta$ meson 
with quarks Q assumed in Figs. 1,2,3. Is it justified ? The second point 
deals with the off-shell (virtually light mass $k^{2} = m_{\eta}^{2}$)
Higgs decay amplitude into two pions.

1- The justification for the $\eta$-quarks coupling 
can be traced back to its Goldstone nature,
to the partially conserved axial current ( PCAC ) and its consequence : 
the Goldberger-Treiman ( GT ) relation. Its well known application is the
$ \eta, \pi \rightarrow  2 \gamma$ rate that gives the number of colors to
be three.

2- The virtually light ($ k^{2} = m_{\eta}^{2}$ ) 
Higgs boson coupling to two
pions can be reliably estimated from the so-called conformal anomaly i.e. 
the trace of the energy- momentum tensor in QCD\cite{R8} 
$\Theta^{\mu}_{\mu} = - \beta_{0} { \alpha_s \over 8 \pi } 
G_{\mu\nu} G^{\mu\nu} $
$(\beta_{0} = 9 $ is the first coefficient of the QCD $\beta$ function). 
The crucial point-as explained in \cite{R9} - 
is that the matrix element of the operator
$\alpha_s G_{\mu\nu} G^{\mu\nu} $
between the two-pion state and the vacuum is nonvanishing in the chiral limit, 
it even does not depend on $\alpha_{s}$ ; 
the (virtually light $k^{2}$) Higgs decay amplitude 
into two pions is found to be \cite{R9} :
\begin{equation}
f_{H\pi\pi}(k^2) = - {g \over \beta_{0}} \hspace{2mm}
{k^2 + 5.5 \hspace{2mm} m_{\pi}^2 \over M_{W} } \label{eq:f}
\end{equation}
where $ g = e /sin\theta_{W}  $ is the standard SU(2) gauge coupling 
which enters also
in the three other vertices of Figs. 1,2,3.
Putting altogether the ingredients, we obtain for the 
$\eta \rightarrow \pi +\pi$ 
decay amplitude the following result :
\begin{equation}
A_{\eta \pi^{+} \pi^{-}} = A_{\eta\pi^0\pi^0} = 
{1 \over 6 \sqrt{3}} \hspace{2mm}
\left( {G_F M_W^2 \over 4 \pi^2} \right)^2 \hspace{2mm}
{m_{\eta}^2 \over M_W^2} \hspace{2mm}
{ m_{\eta}^2 + 5.5 m_{\pi}^2 \over  m_{H}^2 - m_{\eta}^2  } \hspace{2mm}
{X \over f_{\eta}} \hspace{2mm}
\left( 1 + {D_t + E \over C} \right) \label{eq:8}
\end{equation}
with
\begin{equation}
X = m_s^2 (\sqrt{2} cos\theta_P + sin\theta_P) - (m_u^2 + m_d^2) 
\left( { cos\theta_P \over \sqrt{2}}  - sin\theta_P \right) 
\label{eq:9} 
\end{equation}
from which :
\begin{equation}
\Gamma(\eta \rightarrow \pi^{+}\pi^{-}) = 
2 \hspace{2mm} \Gamma(\eta \rightarrow \pi^0 \pi^0) =
{|A_{\eta\pi\pi}|^2 \over 16 \pi m_{\eta}} \hspace{2mm}
\sqrt{1 -{4 m_{\pi}^2 \over m_{\eta}^2} } \label{eq:10}
\end{equation}
 
In Eqs. (14) -(15), the GT like relation 
$g_{\eta Q \overline{Q}}  = m_{Q} / f_{\eta}$
is used, quark color indices are summed up, 
and  $\theta_{P} \simeq -19^{o}$  is
the flavour SU(3) $ \eta -\eta^{'} $ mixing angle 
determined from their two photon rates. 
We take  $f_{\eta} = f_{\pi} \simeq 93$ $ MeV$. 
It turns out that the numerical values of the quantity  
$ Y \equiv  X / f_{\eta} $ 
entering in Eq. (14) are relatively insensitive to the choices of quark
masses : for the constituent ones 
$ m_{s} = 500$ $ MeV$, $m_{u} = m_{d} = 300$ $ MeV$,
we have $Y = 0.79$ $ GeV $ ; for the current ones
$m_{s} = 200$ $ MeV$, $m_{u} = m_{d} = 8$ $ MeV$,
we get $Y = 0.44 $ $GeV$. 
With the constituent mass choice,
we obtain : 
\begin{equation}
Br(\eta \rightarrow \pi + \pi ) =  3.6 \hspace{1mm}\cdot\hspace{1mm} 10^{-26} 
\hspace{2mm}( {100 \hspace{1mm}  GeV  \over M_H } )^{4} 
\end{equation}
such that for the Higgs mass
between 100 $GeV$ and 600 $GeV$, 
the branching ratio into both charged and neutral pions
of the $\eta$ meson  varies in the range 
$ 3.6 \cdot 10^{-26}  -  2.4 \cdot 10^{-29} $,
which is similar although somewhat larger 
than the Jarlskog and Shabalin (JS) \cite{R5} result 
(for the Higgs mass $\leq 250GeV$ ).

Therefore the standard model predicts that existing as well as future 
$\eta$ factories (Saturne, Celsius, Daphne ) could not detect the 
$ \eta \rightarrow \pi +\pi $ mode (unless the Higgs mass is improbably as 
light as 550 $MeV$), implying that {\it unconventional} CP violation 
mechanisms are the only ones that could give rise to its eventual observation.
This fact is not as negative as it seems, since as noted by JS, new CP 
violation mechanisms, what ever they may be, will have a golden opportunity 
to show up in the $ \eta \rightarrow \pi +\pi$ decay. {\it Its eventual 
observation in $\eta$ factories would definitely rule out the standard 
CP violation mechanism (like KM or our mixing), its experimental search is 
even more interesting for this reason}.

In conclusion, we point out that CP violation in $\eta$ decay can occur, 
{\it not by explicit CP non conservation term put by hand in the 
lagrangian, but by its mixing with the Higgs boson through quantum effect}.
It is interesting on its own right.
Conceptually the phenomenon may share a common point (but not to get confused)
with the ABJ anomaly in chiral lagrangian : the chiral symmetry with massless 
$u, d$ quarks has a conserved axial current,
however due to quantum loop effect, the latter is no more conserved.
Also is the induced $\theta$ term with strong CP odd in $QCD$.
In our case, the KM phase is not needed, the interaction is CP conserving,
however, induced by quantum loop, a mixing between {\it states with 
opposite parities} can occur and trigger a CP violating effect, 
in which an unexpected new role of the Higgs boson is revealed.

\vspace{1mm}
\hspace{1cm} \large{} {\bf Acknowledgements}    \vspace{0.1cm}
\normalsize

One of us (X. Y. P.) would like to thank Q. Ho-Kim 
for his critical discussions.

\newpage


\newpage

\large
Figure Captions  :  \normalsize 

\begin{enumerate}
\item
Figure 1 : $\eta $ - Higgs  mixing by two-loop 
$WW$, $\chi\chi$, and $W\chi$ exchange. 
$U_{i}$ denotes $u,c,t$ quarks, 
only the charm quark contribution is important.

\item
Figure 2 : $\eta $ - Higgs  mixing by two-loop 
$U_{i}$ quark exchange via $W$ and $\chi$.
$U_{i}$ denotes $u,c,t$ quarks, 
only the top quark contribution is important.

Similar diagrams with $Z^0$, $\chi^3$ replacing $W$ and $\chi^{\pm}$ 
turn out to be negligible.    

\item
Figure 3 : $\eta $ - Higgs  mixing by two-loop 
via $W$ and $\chi$ exchange.
$U_{i}$ denotes $u,c,t$ quarks. 

Similar diagrams with $Z^0$, $\chi^3$ replacing $W$ and $\chi^{\pm}$
with $U_{i}$ = $s$.

\end{enumerate}
\end{document}